\documentclass[aps,pra,showpacs,superscriptaddress,twocolumn,longbibliography]{revtex4-1}
\usepackage{amsfonts}
\usepackage{subfigure}
\usepackage{amsmath}
\usepackage{txfonts}
\usepackage{amssymb}
\usepackage{amsbsy}
\usepackage{epsfig}
\usepackage{graphicx}
\usepackage{epstopdf}

\begin{document}

\title{The quantum phase transitions of dimer chain driven by an imaginary ac
field}
\author{C. S. Liu}
\date{\today }
\email{csliu@ysu.edu.cn}
\affiliation{Hebei Key Laboratory of Microstructural Material Physics,
School of Science, Yanshan University, Qinhuangdao, 066004, China}

\begin{abstract}
A topologically equivalent tight binding model is proposed to study the quantum phase transitions of dimer chain driven by an imaginary ac field. I demonstrate
how the partner Hamiltonian is constructed by a similarity
transformation to fulfil the $\mathcal{PT}$ symmetry. The $\mathcal{PT}$
symmetry of the partner model allows us to study the topological properties of
the original non-Hermitian model as the Bloch bands of the Hermitian system.
The quantum phase transitions are discussed in different frequency regime.
The approach has the potential applications to investigate the topological
states of matter driven by the complex external parameters.
\end{abstract}

\maketitle

\section{Introduction}

As pioneered by Yang and Lee in 1952, the quantum phase transitions (QPTs)
can also be driven by the complex external parameters \cite{PhysRev.87.404,
PhysRev.87.410}. It is believed that the introduction of the complex
parameters at least gives a deep understanding QPTs \cite{Matveev_2008}. The
interest to study the issue is evoked recently by the realization of the static
complex Zeeman field experimentally in non-Hermitian atomic systems by laser
assisted spin-selective dissipations \cite{Lapp_2019, Nat_commun_855_1}. The application of real ac fields has become a very promising tool to synthesize
novel topological phase termed as Floquet topological
engineering which otherwise would be impossible to achieve in the
un-driven case \cite{PhysRevLett.108.043003, PhysRevLett.111.136402, PhysRevLett.120.243602, PhysRevLett.121.036402, PhysRevLett.124.057001, PhysRevB.95.205125, PhysRevB.97.245430}. A
natural question that arises in this topics is what controllable characters of imaginary ac fields
can bring to non-Hermitian systems although this kinds of field hasn't generated yet.

For the
time-dependent Hamiltonian in the dipolar approximation, the relation
between un-driven and driven system is given by the minimal coupling $%
\mathbf{k\rightarrow k+A}\left( t\right) $ where $\mathbf{A}\left( t\right) $
is the vector potential. One can obtain the time-dependent tight binding
(TB) model with the time-dependent hopping $\tau \left( t\right) _{j,l}=\tau
_{j,l}e^{i\mathbf{A}\left( t\right) \cdot \left( \mathbf{R}_{j}-\mathbf{R}%
_{l}\right) }$ in the inverse Fourier transform. However, for the case of
imaginary ac field $\mathbf{A}\left( t\right) =i\mathbf{A}_{0}\left( t\right) $,
$\tau \left( t\right) _{j,l}=\tau _{j,l}e^{-\mathbf{A}_{0}\left( t\right)
\cdot \left( \mathbf{R}_{j}-\mathbf{R}_{l}\right) }$ and $\tau \left(
t\right) _{j,l}\neq \tau \left( t\right) _{l,j}$ cause the imbalance hopping and non-Hermitianity of the
Hamiltonian.
which is contrast to the real ac field case $\tau \left( t\right) _{j,l}=
\tau \left( t\right) _{l,j}$.

Generally, near the QPTs point, the external real parameters of the
Hamiltonian drive the energy-levels crossing or avoided energy-level
crossing between the ground state and the excited state. The global
curvature in the parameter space can be captured by the geometric phase. Comparing the traditional analysis resorting to the order
parameter and symmetry breaking within the Landau-Ginzburg paradigm, the
geometric phases of the ground state provide a comprehensive reflection of
the ground state characteristics in many-body systems.
However, the non-Hermitian Hamiltonian exists the complex band structure
and exhibits intriguing features with no counterpart in Hermitian cases \cite%
{Rotter_2009, PhysRevB.95.184306, PhysRevE.69.056216, PhysRevLett.118.093002}
except for a non-Hermitian Hamiltonian having simultaneous parity-time ($%
\mathcal{PT}$) symmetry \cite{PhysRevLett.80.5243, PhysRevA.90.012103,
PhysRevA.90.012103}. The imbalance hooping induces non-Hermitian skin effect which causes all
eigenstates are localized exponentially to boundaries, regardless of
topological edge states and bulk states \cite{PhysRevLett.121.086803, PhysRevLett.121.136802, PhysRevB.99.201103, PhysRevLett.124.056802, PhysRevResearch.1.023013, 2020NatPh..16..761X, PhysRevResearch.2.023265, 2020NatPh..16..747H}.
In particular, the bulk-boundary correspondence is
found recently to be breakdown completely for some non-Hermitian systems due to the skin effect
\cite{PhysRevLett.116.133903, Xiong_2018, PhysRevLett.121.026808,
MartinezAlvarez2018, PhysRevB.102.041119, PhysRevB.101.045415, PhysRevB.102.041119, 2020NatPh..16..761X}.

Concerning the non-Hermitian topological phases, the general Brillouin
zone is used to understand the bulk-boundary correspondence of non-Hermitian system where a complex-valued wave vector
is introduced to capture unique feature of non-Hermitian bands \cite%
{PhysRevLett.121.086803, 2019arXiv190210958Y, 2019arXiv190502211S,
PhysRevB.100.035102}. The real part of the wave vector is from the
periodicity of the system according to the Bloch theorem. The imaginary part
of complex-valued wave vector is responsible for non-Hermitian skin effect.
Experimentally, the non-Hermitian bulk-boundary correspondence has been demonstrated in discrete-time non-unitary quantum-walk dynamics of single photons \cite{2020NatPh..16..761X} and in topolectrical circuits \cite{2020NatPh..16..747H}.

Periodically driven non-Hermitian systems involve not only the non-Hermitian topological phases, but also Floquet topological
phases. It has exhibit rich topological phases and non-Hermitian
skin effect, without analogs in their static or Hermitian counterparts \cite{PhysRevLett.123.190403, PhysRevB.101.045415, PhysRevB.102.041119, PhysRevB.98.205417, 2020arXiv200713499C, PhysRevB.101.014306}. In particular, some of the studies focus on the periodic quenching a
Hamiltonian from $H_1$ for
the first half period to $H_1$ for the second one \cite{PhysRevB.101.045415, PhysRevB.102.041119, PhysRevB.98.205417}. The key role played
by periodic driving is changing symmetry and inducing an
effective long-range hopping in lattice systems \cite{PhysRevB.87.201109}.

Controllable gain and loss of the available experimental setups
make concrete realizations of non-Hermitian lattice model
possible, such as a non-Hermitian version of the topological
SSH model \cite{PhysRevLett.115.040402, Weimann2016Topologically, 2019arXiv190711619G, Schomerus:13}. However, the non-Hermitianity induced by an imaginary ac
field is not involved.

Motivated by the above considerations, I investigate the imaginary parameter driven QPTs by the analysis of an
ac-driven dimer chain.
I introduce a partner model without non-Hermitian
skin effect that shares the same topological phase diagrams \cite%
{CPB.29.10302}. The partner of non-Hermitian Hamiltonian can be constructed
by adjusting the imbalance hopping to fulfilling inverse or reflection
symmetry. I demonstrate that partner Hamiltonian can be obtained by a
similarity transformation and prove their topologically equivalent. In
particular, the partner Hamiltonian have a $\mathcal{PT}$ symmetry which
allows to study the topological properties as the Bloch bands of the
Hermitian system.

The merit of this method is that the bulk boundary
correspondence is still effective without the introduction of the
generalized bulk boundary correspondence and non-Bloch winding number. As
shown below, the imaginary ac field drives the initial Bloch band splits into
Floquet-Bloch bands. Interestingly, the ac-driven band inversion will lead
to interesting topological states of matter which otherwise would be
inaccessible in the real field case.

The remainder of this paper is organized as follows. In Sec. \ref{Model}, I
present the 2D effective TB Hamiltonian of SSH model driven by an imaginary ac
electric fields. I show how the amplitude of the vector potential controls
the renormalization of the system parameters and leads to the
non-Hermitianity of the effective Hamiltonian. In subsection \ref{The
partner model}, I present a partner model of the original model and
illuminate that the original and modified models are related by a similarity
transformation. Due to the topologically equivalence of the two model, I
study the QPTs in low and intermediate frequency regime with the partner
model in subsection \ref{Topological invariant and phase transition in low
frequency limit} and \ref{Beyond the low and high frequency limit}. Finally,
I present a summary and discussion in Sec. \ref{Summary}.

\section{Model}

\label{Model}

I consider the one-dimensional SSH model \cite{PhysRevLett.42.1698} driven
by an imaginary ac electric fields. The imaginary ac electric field described by an
imaginary vector potential $iA(t)=iA_{0}\sin (\omega t)$ of frequency $\omega $
is applied in the chain direction. With the standard Peierls substitution $%
k\rightarrow k+A$, the Hamiltonian is given by $H(k,t)=\Psi _{k}^{\dag
}\left( t\right) h(k,t)\Psi _{k}\left( t\right) $ where $\Psi _{k}^{\dag
}\left( t\right) =\left[ c_{k,A}^{\dagger }\left( t\right) ,c_{k,B}^{\dag
}\left( t\right) \right] $ and the non-Hermitian operator
\begin{equation}
h(k,t)=\left(
\begin{array}{cc}
0 & \tau +\tau ^{\prime }e^{A\left( t\right) }e^{-ik} \\
\tau +\tau ^{\prime }e^{-A\left( t\right) }e^{ik} & 0%
\end{array}%
\right) .  \label{Hamiltonian}
\end{equation}%
$\tau $ and $\tau ^{\prime }$ are the two hopping parameters of the dimer
chain.

The model can be investigated by the Floquet-Bloch ansatz \cite{PR304p229,
PhysRevLett.110.200403}. The essence of this method is mapping the
time-dependent Schr\"{o}dinger equation to the eigen problem
\begin{equation}
\bar{h}(k,t)\Psi _{k}\left( t\right) =\Psi _{k}\left( t\right) \epsilon _{k},
\label{HF}
\end{equation}%
here
\begin{equation*}
\bar{h}(k,t)=h(k,t)-i\partial _{t}=\left(
\begin{array}{cc}
-i\partial _{t} & \tau +\tau ^{\prime }e^{A\left( t\right) }e^{-ik} \\
\tau +\tau ^{\prime }e^{-A\left( t\right) }e^{ik} & -i\partial _{t}%
\end{array}%
\right)
\end{equation*}%
is defined as the Floquet operator, and $\epsilon _{k}=\left(
\begin{array}{cc}
\epsilon _{1,k} & 0 \\
0 & \epsilon _{2,k}%
\end{array}%
\right) $ is the quasienergy. The quasi-energies $\epsilon _{k}$ and the
Floquet states $\Psi _{k}\left( 0\right) $ can be obtained by diagonalizing
the one-period propagator $U(T)$. The propagator $U(T)$ can be solved by
integrating the following evolution equation numerically,
\begin{equation*}
i\frac{\partial }{\partial t}U(t)=h(t)U(t).
\end{equation*}%
in one period with the initial condition $U(0)=I_{0}$, here $I_{0}$ is the
unit matrix.

\begin{figure}[tbp]
\begin{center}
\includegraphics[width=9cm]{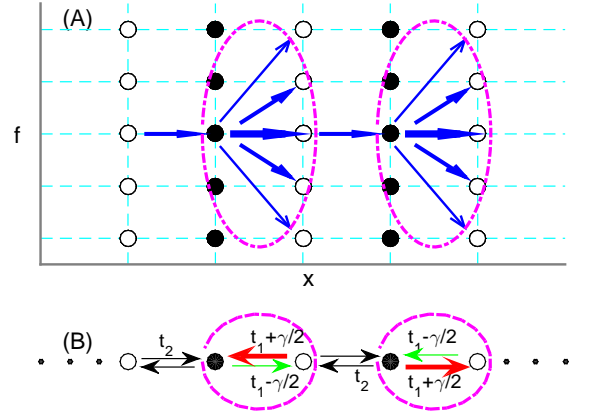}
\end{center}
\caption{(A) The effective 2D lattice of the periodically driven dimer
chain. The additional ${f}$ dimension is spanned by $\{ n\}$ due to the time
periodicity. The width of the arrows shows the hopping intensity
qualitatively. (B) The partner model in the $x$ direction. Adjusting the
imbalance hopping between the adjacent unit cells (the magenta-dotted
ellipse boxes) leads to the non-Hermitian skin effect disappearing in the $x$
direction. }
\label{fig1}
\end{figure}

Alternatively, the driven dimer chain is mapped to time independent 2D
lattices. The quasienergy can be obtained by diagonlizating the effective
Hamiltonian \cite{PhysRevLett.110.200403}. This method will be used in the
following discussions and summarized briefly here. Using the Fourier
transformation $\Psi _{k}\left( t\right) =\sum_{n}\Phi _{k,n}e^{in\omega t}$
and with the inner product $\langle \langle \cdot \cdot \cdot \rangle
\rangle =\frac{1}{T}\int_{0}^{T}\langle \cdot \cdot \cdot \rangle dt$, the
Floquet operator\ is transformed to
\begin{equation*}
\bar{H}(k,k_{f})=\sum_{m,n}\Phi _{k,m}^{\dag }\bar{h}_{m,n}\Phi _{k,n}
\end{equation*}%
\begin{equation}
\bar{h}_{m,n}=\left(
\begin{array}{cc}
-n\omega \delta _{n,m} & \tau \delta _{n,m}+B_{m,n} \\
\tau \delta _{m,n}+B_{n,m} & -n\omega \delta _{n,m}%
\end{array}%
\right)  \label{Floquet operator}
\end{equation}%
where $\Phi _{k,\alpha }^{\dag }=\left(
\begin{array}{cc}
c_{k,A,\alpha }^{\dag } & c_{k,B,\alpha }^{\dag }%
\end{array}%
\right) \ $with $\alpha =m,n$ and $B_{m,n}=\tau ^{\prime
}e^{-ik}J_{m-n}\left( -iA_{0}\right) $, $B_{n,m}=\tau ^{\prime
}e^{ik}J_{n-m}\left( iA_{0}\right) $. $J_{\nu }\ $is the $\nu $th Bessel
function of the first kind. The diagonal term $-n\omega \delta _{n,m}$ of
Eq. (\ref{Floquet operator}) is equivalent to an effective electric field $%
\omega $ in $f$ direction and break the space inversion symmetry. Eq. (\ref%
{Floquet operator}) is an infinite matrix because $n,m$ are integer. Shown
in Fig. \ref{fig1}(A), the one-dimensional ac-driven model is exactly mapped
to two dimensional TB problem in the composed Hilbert $\mathcal{S}=\mathcal{H%
}\otimes \mathcal{T}$, where $\mathcal{H}$ is the ordinary Hilbert space and
$\mathcal{T}$ is the space of T-periodical function.

In the low frequency regime $\omega \ll \tau ,\tau ^{\prime }$, the diagonal
term $-n\delta _{n,m}$ in Eq. (\ref{Floquet operator}) can be neglected. The
Floquet operator is an effective TB model in the frequency space $\mathcal{T}
$. The TB model can be transformed to the corresponding momentum ${k_{f}}$\
space:
\begin{equation}
\hat{h}\left( k,k_{f}\right) \mathbb{=}\tau \left(
\begin{array}{cc}
0 & t_{2}+t_{-}e^{-ika} \\
t_{2}+t_{+}e^{ika} & 0%
\end{array}%
\right) ,  \label{the Floquet operator in 2D space}
\end{equation}%
here
\begin{eqnarray}
t_{1,\pm }\left( k_{f}\right) &=&I_{0}\pm \sum_{l=2n+1}2\left( -1\right)
^{n}I_{l}\sin lk_{f}\pm \sum_{l=2n}2\left( -1\right) ^{n}I_{l}\cos lk_{f}
\notag \\
&=&t_{1}\pm \frac{\gamma }{2}.  \label{the imbalance hopping terms}
\end{eqnarray}%
Shown in Fig. \ref{fig1}(B), the TB model of the Floquet operator in Eq. (%
\ref{the Floquet operator in 2D space}) is dimer chain with the imbalance
hopping $t_{1}+\frac{\gamma }{2}$ in the left and $t_{1}-\frac{\gamma }{2}$
right directions. I have used the relationship $J_{-\nu }\left( A_{0}\right)
=\left( -1\right) ^{\nu }J_{\nu }\left( A_{0}\right) $ and $J_{\nu }\left(
iA_{0}\right) =i^{\nu }I_{\nu }\left( A_{0}\right) $, $I_{\nu }\ $is the $%
\nu $th modified Bessel function of the first kind \cite{Table_of_Integrals}%
. In fact, the effective SSH model can also be written as
\begin{equation}
\hat{h}\left( k,k_{f}\right) \mathbb{=}\tau \left(
\begin{array}{cc}
0 & t_{+}+t_{2}e^{-ika} \\
t_{-}+t_{2}e^{ika} & 0%
\end{array}%
\right) .  \label{BA lattice}
\end{equation}%
The former is AB lattice and the latter is the BA lattice. They have the
same phase transition point. The following discussions are focused on the
equivalent BA lattice.

Out of the low frequency regime, we must diagonalize the one-period
propagator $U(T)$ or the full Floquet matrix (\ref{Floquet operator}) to get
the quasienergy due to the Floquet bands coupling. In the high frequency
regime $\omega \gg \tau ,\tau ^{\prime }$ however, the diagonal terms in Eq.
(\ref{Floquet operator}) are important. The quasienergy bands have a large
band gap $\omega $ and the matrix is approximately block diagonal. The
topological phase transitions occur in the lowest Floquet band $m=n=0$. In
this case, the effective Hamiltonian in Eq. (\ref{Floquet operator}) is
reduced to a the $2\times 2$ hermitian matrix:
\begin{equation}
\bar{h}_{k}=\tau \left(
\begin{array}{cc}
0 & I_{0}+t_{2}e^{-ik} \\
I_{0}+t_{2}e^{ik} & 0%
\end{array}%
\right) .  \notag
\end{equation}
It is the SSH model and the topological phase transition occurs at $%
t_{2}=\tau ^{\prime }/\tau =I_{0}$. This is different from the real ac field
case where the topological phase transition occurs at $t_{2}=J_{0}$ \cite%
{PhysRevLett.110.200403}.

\section{Results}

\subsection{The partner model}

\label{The partner model}

The idea of constructing the partner Hamiltonian is from that the
topological boundary states are protected by the symmetry of system and are
immune to perturbations. It is reasonable to speculate that, for a class of
non-Hermitian systems with non-Hermitian skin effect, there is a partner
without non-Hermitian skin effect that has the same symmetry. As such, the
topological invariants of the original models with non-Hermitian skin effect
can be obtained from their partners, which share topological phase diagrams
can be calculated in an easier way.

I construct the partner of the Floquet operator shown in Fig. \ref{fig1}(B).
The reflection symmetry of the model is due to the change of the hopping
terms $t_{1,\pm } $ alternately in the adjacent unit cell, and causes the
skin effect to disappear \cite{PhysRevB.99.125103}. For simplicity of the
introduction, the diagonal term $-n\omega \delta _{n,m}$ in Eq. (\ref%
{Floquet operator}) is removed firstly and $k_{f}$ is good quantum number.
As shown below, the existence of the diagonal term don't modify the
conclusion. The partner Hamiltonian reads $\check{H}\left( k,k_{f}\right)
=\Omega _{k}^{\dag }\check{h}\Omega _{k}$ with
\begin{equation}
\check{h}=\left(
\begin{array}{cccc}
0 & t_{1,+} & 0 & t_{2}e^{-ik} \\
t_{1,-} & 0 & t_{2} & 0 \\
0 & t_{2} & 0 & t_{1,-} \\
t_{2}e^{ik} & 0 & t_{1,+} & 0%
\end{array}%
\right)  \label{the partner Hamiltonian}
\end{equation}%
where $\Omega _{k}^{\dag }\left( k_{f}\right) =\left(
\begin{array}{cccc}
a_{k}^{\dag }, & b_{k}^{\dag }, & c_{k}^{\dag }, & d_{k}^{\dag }%
\end{array}%
\right) $ is the creation operation of the lattice (abcd) in unit cell of
Fig \ref{fig1}. The partner Hamiltonian $\check{h}$ has the $\mathcal{PT}$
symmetry and follows the relation $[\mathcal{PT},\check{h}]=0$. $\mathcal{P}$
and $\mathcal{T}$ are defined as the space-reflection (parity) operator and
the time-reversal operator, whose effects are given by $k\rightarrow -k$, $%
x\rightarrow -x$ and $k\rightarrow -k$, $x\rightarrow x$, $i\rightarrow -i$,
respectively. The wave vectors $k$ and $k_{f}$ are real.

The partner model and original model are topologically equivalent. If we
introduce a parameter $\theta $ in the model, i.e.%
\begin{equation}
h\left( \theta \right) =\left(
\begin{array}{cccc}
0 & t_{1}+\frac{\gamma }{2} & 0 & t_{2}e^{-ik} \\
t_{1}-\frac{\gamma }{2} & 0 & t_{2} & 0 \\
0 & t_{2} & 0 & t_{1}-\frac{\gamma }{2}\cos \theta \\
t_{2}e^{ik} & 0 & t_{1}+\frac{\gamma }{2}\cos \theta & 0%
\end{array}%
\right) ,  \label{topologically equivalent}
\end{equation}%
the original Hamiltonian $\hat{h}$ of Eq. (\ref{the imbalance hopping terms}%
) can be continuously deformed into the partner Hamiltonian $\check{h}$ of
Eq. (\ref{the partner Hamiltonian}) when changing the parameter $\theta $
from $\pi $ to $0$. The Hamiltonian in Eq. (\ref{topologically equivalent})
also has a chiral symmetry $\Sigma _{z}^{-1}h\left( \theta \right) \Sigma
_{z}=-h\left( \theta \right) $ with $\Sigma _{z}=\sigma _{0}\otimes \sigma
_{z}$ and $\sigma _{0}\ $and $\sigma _{z}$ are unity matrix and the Pauli
matrices respectively. The chiral symmetry ensures that the eigenvalues of
Hamiltonian $h\left( \theta \right) $ appear in $\left( \varepsilon
_{j},-\varepsilon _{j}\right) $ pairs with $j=1,2$.

In the real-space, the original Hamiltonian $\hat{h}$ and partner
Hamiltonian $\check{h}$ are related by a similarity transformation
\begin{equation}
\check{h}=S_{0}^{-1}\hat{h}S_{0}.
\label{relationship betwen original and partner}
\end{equation}
$S_{0}$ is a diagonal matrix whose diagonal elements are $%
\{1,r^{2},r^{2},r^{2},\cdots ,r^{L/4},r^{L/4+1},r^{L/4+1},r^{L/4+1}\}$ and $%
r=\sqrt{\left\vert \frac{t_{1}-\gamma /2}{t_{1}+\gamma /2}\right\vert }$
here $L$ is the number of unit cell. The real-space eigen-equation $\hat{h}%
|\psi \rangle =E|\psi \rangle $ is equivalent to $\check{h}|\bar{\psi}%
\rangle =E|\bar{\psi}\rangle $ with $|\bar{\psi}\rangle =S_{0}^{-1}|\psi
\rangle $ where the eigenvalue $E$ remains unchanged.

The diagonal matrix $S_{0}$ can be decomposed into the products of $S_{1}$
and $S_{2}$ ($S_{0}=S_{1}S_{2}$) where $S_{1}$ is a diagonal matrix whose
diagonal elements are $\{1,r,r,r^{2},r^{2},\cdots ,r^{L-1},r^{L-1},r^{L}\}$
and $S_{2}$ is a diagonal matrix whose diagonal elements are $%
\{1,r,r,1,...,1,r,r,1\}$. The modified Hamiltonian $\check{h}$ can be
constructed by two similarity transformations. With the similarity
transformation as done in Eq.
\begin{equation}
\tilde{h}=S_{1}^{-1}\hat{h}S_{1},  \label{similarity transformation in PRL}
\end{equation}%
$\hat{h}$ becomes the standard SSH model
\begin{equation*}
\tilde{h}=(\tilde{t}_{1}+t_{2}\cos k)\sigma _{x}+t_{2}\sin k\sigma _{y}
\end{equation*}%
for $|t_{1}|>|\gamma /2|$, with intracell and intercell hopping $\tilde{t}%
_{1}=\sqrt{(t_{1}-\gamma /2)(t_{1}+\gamma /2)}\ $and$\quad t_{2}=\tau
^{\prime }$. This result has been obtained in Ref. \cite%
{PhysRevLett.121.086803}. Then doing the other similarity transformation $%
\check{h}=S_{2}^{-1}\tilde{h}S_{2}$, $\tilde{h}$ becomes the partner
Hamiltonian $\check{h}$ in Eq. (\ref{the partner Hamiltonian}).

The corresponding eigenvalues $\hat{\varepsilon}$ of the original operator $%
\hat{h}$ are complex and the corresponding eigenvalues $\check{\varepsilon}$
of the partner operator $\check{h}$ are real. The relationship between $\hat{%
\varepsilon}$ and $\check{\varepsilon}$ can be understood as follow. The
eigenvalues $\hat{\varepsilon}$ can be obtained by a unitary transformation,
i.e. $\hat{\varepsilon}=\hat{U}_{L}^{\dag }\hat{h}\hat{U}_{R}$, here $\hat{U}%
_{L}^{\dag }\hat{U}_{R}=\hat{U}_{R}^{\dag }\hat{U}_{L}=I_{0}$ and $I_{0}$ is
the\ unity matrix. Due to the non-Hermitianity of operator $\hat{h}$, the
unitary operator $\hat{U}_{L}$ and $\hat{U}_{R}$ must be differentiated. The
corresponding eigenvalues $\check{\varepsilon}$ of the partner operator $%
\check{h}$ can be obtained by the other unitary transformation $\check{%
\varepsilon}=\check{U}^{\dag }\check{h}\check{U}$ and $\check{U}^{\dag }%
\check{U}=\check{U}\check{U}^{\dag }=I_{0}$. Due to the $\mathcal{PT}$
symmetry of $\check{h}$, it is unnecessary to differentiate the left and
right unity matrixs. From the relationship between the original and partner
operators in Eq. (\ref{relationship betwen original and partner}), I can
easy get the relationship of eigenvalue matrix between original and partner
operators $\hat{\varepsilon}=U_{L}^{\dag }\check{\varepsilon}U_{R}$. It is
also a unitary transformation $U_{L}^{\dag }U_{R}=U_{R}^{\dag }U_{L}=I_{0}$
here the left and right unity matrixes must be distinguished with $%
U_{L}^{\dag }=\hat{U}_{L}^{\dag }S_{0}\check{U}$ and $U_{R}=\check{U}^{\dag
}S_{0}^{-1}\hat{U}_{R}$.

It should point that this method is general effective to the model including
the nearest-neighbor interaction only. When the next nearest-neighbor term $%
t_{3}$ exists in the model further, this method is general fail since it is
difficult to guarantee the reflection symmetry of the $t_{2}$ and $t_{3}$
terms simultaneously under the similarity transformation.

When the diagonal term $-n\omega \delta _{n,m}$ in Eq. (\ref{Floquet
operator}) is included in the model, we can also use this method to construct the
partner model. The reason is that the similarity transformation of Eq. (\ref{similarity
transformation in PRL}) is effectively applied to the $x$ direction and the
diagonal part $-n\omega \delta _{n,m}$ of matrix (\ref{Floquet operator})
remains unchanged since it is equivalence to electric field applying to the $%
f$ direction. The $\mathcal{PT}$ transformation also occurs in the $x$
direction, the partner Hamiltonian also have the $\mathcal{PT}$ symmetry and
its eigenvalues are real. When changing the parameter $\theta $ from $\pi $
to $0$, the original Hamiltonian $\bar{h}$ of Eq. (\ref{Floquet operator})
can be continuously deformed into the partner Hamiltonian and also remains $%
-n\omega \delta _{n,m}$ unchanged. So the partner is a topological
equivalent to the original model.

\subsection{Topological invariant and phase transition in low frequency limit%
}

\label{Topological invariant and phase transition in low frequency limit}

From Eq. (\ref{the imbalance hopping terms}), the external ac field modifies
the effective hopping of the model which changes the QPTs. $A_{0}$ and $k_{f}$ are adjustable parameters of the one
dimensional problem. The Hamiltonian (\ref{the Floquet operator in 2D space}%
) and its partner (\ref{the partner Hamiltonian}) belongs to the BDI class,
the winding number along the $k$ direction is the topological invariant that
differentiates the system from an ordinary insulator \cite{Bernard_2002,
PhysRevLett.110.200403}. To calculate the winding number of the partner
Hamiltonian $\check{h}_{k}$ in Eq. (\ref{the partner Hamiltonian}), the
Hamiltonian $\check{h}_{k}$ is transformed into block off-diagonal form
\begin{equation}
U\check{h}_{k}U^{-1}=\left(
\begin{array}{cccc}
0 & 0 & t_{1}+\frac{1}{2}\gamma & t_{2}e^{ik} \\
0 & 0 & t_{2} & t_{1}-\frac{1}{2}\gamma \\
t_{1}-\frac{1}{2}\gamma & t_{2} & 0 & 0 \\
t_{2}e^{-ik} & t_{1}+\frac{1}{2}\gamma & 0 & 0%
\end{array}%
\right)  \label{block off-diagonal Hamiltonian}
\end{equation}%
with the unitary matrix%
\begin{equation*}
U=\left(
\begin{array}{cccc}
0 & 0 & 0 & 1 \\
0 & 1 & 0 & 0 \\
0 & 0 & 1 & 0 \\
1 & 0 & 0 & 0%
\end{array}%
\right) .
\end{equation*}%
With the block off-diagonal Hamiltonian in Eq. (\ref{block off-diagonal
Hamiltonian}), the winding number is defined by \cite{PhysRevB.83.085426,
PhysRevX.4.011006}
\begin{eqnarray}
\mathcal{W} &=&-\oint \frac{dk}{2\pi i}\partial _{k}\left\{ \ln \left[
\mathbf{Det}\left(
\begin{array}{cc}
\frac{1}{2}\gamma +t_{1} & t_{2}e^{ik} \\
t_{2} & t_{1}-\frac{1}{2}\gamma%
\end{array}%
\right) \right] \right\}  \label{winding} \\
&=&-\oint \frac{dk}{2\pi i}\partial _{k}\left[ \ln \left( t_{1}^{2}-\frac{%
\gamma ^{2}}{4}+t_{2}^{2}e^{ik}\right) \right] .  \notag
\end{eqnarray}%
The winding vector is $t_{1}^{2}-\frac{\gamma ^{2}}{4}+t_{2}^{2}e^{ik}$. We
get the transition points are%
\begin{equation}
t_{1}^{2}=t_{2}^{2}+\frac{\gamma ^{2}}{4},  \label{transition points}
\end{equation}%
namely, $\tilde{t}_{1}=t_{2}$ which is just the results obtained in Ref.
\cite{PhysRevLett.121.086803}.

\begin{figure}[tbp]
\begin{center}
\includegraphics[width=8cm]{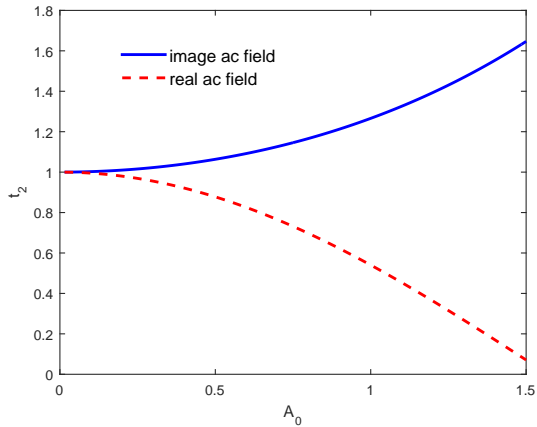}
\end{center}
\caption{The phase boundaries of the effective 2D lattice model driven by
the imaginary ac field (blue solid line) and the real ac field (red dashed
line). The upper part of the boundary is in topological phase and the lower
part of the boundary is in insulator phase.}
\label{fig2}
\end{figure}

In the present example of the driven dimer chain, the topologically phase
boundary is related with the equation (\ref{transition points}). In the
limit $A_{0}\rightarrow 0$, $\left\vert \gamma /2\right\vert \rightarrow 0$
which results to $t_{1}=\tau ^{\prime }/\tau =I\left( A_{0}\right)
\rightarrow 1$. This is the phase transition point of bared SSH model. If we
assume $A_{0}\leq 1.5$, the contributions from $I_{l>6}\left( A_{0}\right) $
can be neglected. With a fixed $A_{0}$, the parameter $\left\vert \gamma
/2\right\vert $ in Eq. (\ref{the imbalance hopping terms}) can be taken in
the range from the minimum $\left\vert \gamma _{\min }/2\right\vert $ to the
maximum $\left\vert \gamma _{\max }/2\right\vert $ when changing $k_{f}$
from $0$ to $2\pi $. So the range of the parameter $\sqrt{t_{1}^{2}-\gamma
^{2}/4}\ $is from $t_{1,\min }=\sqrt{I_{0}^{2}-\gamma _{\max }^{2}/4}$ to $%
t_{1,\max }=\sqrt{I_{0}^{2}-\gamma _{\min }^{2}/4}$. When $t_{2}\geqslant
t_{1,\max }$, the model is in the topological phase. While the case of $%
t_{1,\min }<\left\vert t_{2}\right\vert <t_{1,\max }$, there are some $k_{f}$
with $\left\vert t_{2}\right\vert \geqslant \sqrt{t_{1}^{2}-\gamma ^{2}/4}$
and some $k_{f}$ with $\left\vert t_{2}\right\vert <\sqrt{t_{1}^{2}-\gamma
^{2}/4}$. The former is corresponding to the topological phase and the
latter is corresponding to the topological trivial phase. The model is in
the topological phase, actually. When parameters $\left\vert
t_{2}\right\vert <t_{1,\min }$ however, it is in the trivial case of SSH model.
Therefore, the topological phase transition point is at $\left\vert
t_{2}\right\vert =t_{1,\min }$. $\left\vert t_{2}\right\vert =t_{1,\min }$
as function of $A_{0}$ is plotted in Fig. \ref{fig2} (blue solid line).

It is interesting to compare the phase boundary to the case driven by a real
ac field. When the SSH model driven by a real ac field, the effective
Hamiltonian $\hat{h}\left( k,k_{f}\right)$ is given by
\begin{equation*}
\hat{h}\left( k,k_{f}\right) \mathbb{=}\tau \left(
\begin{array}{cc}
0 & t_{1}+t_{2}e^{-ika} \\
t_{1}^{\ast }+t_{2}e^{ika} & 0%
\end{array}%
\right) ,
\end{equation*}%
here
\begin{eqnarray}
t_{1}\left( k_{f}\right) &=&J_{0}+2i\sum_{l=2n+1}J_{l}\sin
lk_{f}+2\sum_{l=2n}J_{l}\cos lk_{f}  \notag \\
&=&\left\vert t_{1}\right\vert e^{-i\theta }.
\label{the effective hopping of real ac field}
\end{eqnarray}%
Follow the analysis in the case of imaginary ac field, the topological phase
transition point is at $t_{2}=t_{1,\min }$. $t_{2}=t_{1,\min }$ as function
of $A_{0}$ is plotted in Fig. \ref{fig2} (red dashed line).

\subsection{Beyond the low and high frequency limit}

\label{Beyond the low and high frequency limit}

\begin{figure}[tbp]
\begin{center}
\includegraphics[width=9cm]{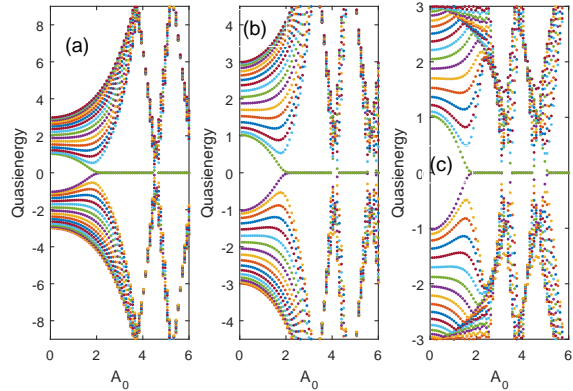}
\end{center}
\caption{Quasienergies $\check{\protect\varepsilon}$ of the equivalent tight
binding model vs $A_0$ for different frequencies $\protect\omega$ under open
boundary conditions. (a) shows the quasi-energies near the high frequency
limit $\protect\omega = 16$, where the bands are not coupled. (b) the
coupling between different Floquet bands leads the bands inversions and open
a gap with the decreasing the frequency at $\protect\omega = 12$. (c) shows
the topological phase transition occurs at the exact crossings between
conduction and valence band with $\protect\omega = 6$.}
\label{fig3}
\end{figure}

I study the partner model numerically to understand the topological phase
beyond the low and high frequency limit. The eigenvalues of the equivalent
model are real due to its $\mathcal{PT}$ symmetry. Since the high level
modified Bessel function compared with that of zero level are small enough
to be neglected, I numerically diagonalize the partner of Eq. (\ref{Floquet
operator}) without need a larger number of sidebands to reach convergence.
In the present case $A_0=4$ and $I_{7}\left( A_0\right) /I_{0}\left( A_0\right) =0.0037$, the matrix
element $I_{m-n>7}$ can be neglected. Fig. \ref{fig3} shows quasienergy
spectrum vs $A_{0}$ for $n,m=7$ sidebands.

When $\omega >18$, the system is in the high frequency regime and the
critical point $A_{0}=1.8079$ is almost fixed and meets $I_{0}\left(
A_{0}\right) =t_{2}=\tau ^{\prime }/\tau =2$. When $A_{0}>1.8079$ in Fig. %
\ref{fig3}(a), the system exists the topologically protected states. With
decreasing the frequency from the high frequency limit, the coupling between
different Floquet bands cannot be neglected and induces the bands inversions
near $A_{0}=4$ and $A_{0}=5.6$ in Fig. \ref{fig3} (b). Bands inversions
result to the opening and closing the band gap and change the topological
properties of the model. The topological phase transition occur at the exact
crossings between conduction and valence band in Fig. \ref{fig3} (c).

\section{Summary}

\label{Summary}

In summary, with the Floquet-Bloch approach, we have mapped the dimer chain
driven by an imaginary ac field to a 2D effective TB model to study the QPTs
driven by an imaginary external parameter. To investigate the QPTs of the
original non-Hermitian model, I construct a topologically equivalent model
which fulfils the $\mathcal{PT}$ symmetry. The merit of the method is the
real energy spectra of the partner model allows to study the topological
properties as the Bloch bands of the Hermitian system without the
introduction of the generalized Brillouin zone. The similarity
transformation is given to illuminate the relationship between the partner
Hamiltonian and the original Hamiltonian. I have obtained the phase boundary
of driven dimer chain in different frequency regime. This method is expected
useful to study the non-Hermitian systems with the non-Hermitian skin
effect. The predicting novel topological states of matter which otherwise
would be inaccessible in the real field case can be realized experimentally.

\begin{acknowledgments}
This work was supported by Hebei Provincial Natural Science Foundation of
China (Grant No. A2012203174, No. A2015203387) and National Natural Science
Foundation of China (Grant No. 10974169, No. 11304270).
\end{acknowledgments}


%

\end{document}